\documentclass[thmsa,onecolumn,a4paper]{article}

\begin{document}

\author{Alain de Roany, Bertrand Chauvineau, Jos\'{e} A.\ de Freitas Pacheco}
\title{Visser's massive gravity bimetric theory revisited}
\date{University of Nice-Sophia Antipolis, Observatoire de la C\^{o}te
d'Azur, Laboratoire Cassiop\'{e}e, BP 4229, Nice Cedex 4, France\\
e-mail : deroany@oca.eu, Bertrand.Chauvineau@oca.eu, pacheco@oca.eu}
\maketitle

\begin{abstract}
A massive gravity theory was proposed by Visser in the late nineties. This
theory, based on a backgroung metric $b_{\alpha \beta}$ and on an usual
dynamical metric $g_{\alpha \beta}$ has the advantage of being free of
ghosts as well as discontinuities present in other massive theories proposed
in the past. In the present investigation, the equations of Visser's theory
are revisited with a particular care on the related conservation laws.\ It
will be shown that a multiplicative factor is missing in the graviton tensor
originally derived by Visser, which has no incidence on the weak field
approach but becomes important in the strong field regime when, for
instance, cosmological applications are considered. In this case, contrary
to some previous claims found in the literature, we conclude that a
non-static background metric is required in order to obtain a solution able
to mimic the $\Lambda$CDM cosmology.
\end{abstract}

\baselineskip12truept

\bigskip

KEY WORDS : massive gravity - bimetric theory - FRW cosmology

\bigskip

\noindent \textbf{I. INTRODUCTION}

\bigskip

Studies of the luminosity distance as a function of the redshift of type Ia
supernovae suggest that the expansion of the universe is presently
accelerated [1,2]. These data as well as those derived from the
space probe WMAP on the cosmic microwave background [3] can be explained
by the inclusion of the so-called "cosmological constant" term in Einstein's
equations. Arguments against this possibility have been raised in the
literature, in particular the "coincidence" problem and the interpretation
of such a term as the vacuum energy density. However, some authors believe
that these objections do not represent real difficulties for the theory
[4]. They claim that the "coincidence" problem is ill defined and that
the identification of the cosmological constant with the vacuum energy
density is probably a mistake!

Independently of the reality or not of these difficulties for the 
$\Lambda$CDM cosmology, alternative models have been proposed in the literature. In a
first class of models, the usual thermodynamic properties of the
constituents of the universe are modified and the acceleration of the
expansion is driven by a negative pressure term associated either to
particle production (see, for instance, [5] and references therein) or to
a bulk viscosity term ([6] and references therein). In other class of
models, the existence of new fields in nature responsible for the
acceleration are postulated (scalar fields in GR) as well as modifications
in the Einstein-Hilbert action ($f(R)$ theories) [7-9] or
scalar-tensor theories [9-17].

In a particular class of theories (massive gravity), the graviton has a
small but non-zero mass. These theories have a long history and present
several difficulties. Fierz and Pauli (FP) noticed that the mass term must
be quadratic for a Lorentz invariant massive spin-2 theory, otherwise a
\textquotedblleft ghost" appears in the spectrum [18]. A major difficulty
with the FP theory is the van Dam-Veltman-Zakharov discontinuity
[19,20]. In other words, in the linearized FP theory the extra scalar
mode of the graviton does not disappear and remains coupled to matter even
in the limit of a vanishing graviton mass and, consequently, in such a limit
the linear equations of general relativity (GR) are not recuperated.
Moreover, the prediction of the quadratic FP theory for the light bending
effect differs from GR, practically ruling out the FP theory. However,
Vainshtein [21] from the analysis of static spherically symmetric
solutions, argued that the linear FP theory is only valid for distances
larger than a certain scale dubbed the Vainshtein radius $R_{V}$, which goes
to infinity as the graviton mass goes to zero. For distances less than 
$R_{V}$, around a static spherically symmetric source of mass $M$, the full
non-linear strongly coupled theory has to be considered in order to recover
GR. In the past years, the problem of the continuous matching between
solutions inside and outside the Vainshtein radius have been extensively
debated in the literature [22-25].

A different approach was proposed by Visser [26], who introduced a
background metric not subjected to any dynamical equation. The mass term in
this theory depends both on the dynamical and on the background metric in
such a way that in the linear limit, the massive field obeys a Klein-Gordon
equation with a source term. GR is recovered when the graviton mass
vanishes. The massive field in Visser's theory has six degrees of freedom:
five spin-2 and one scalar [26]. This theory was applied to cosmology by
different authors [27-30], who claim that the
resulting dynamical equations based on such a theory are able to explain the
present observed acceleration of the universe and to satisfy other
cosmological tests like the distance scale provided by the baryon acoustic
peak and the cosmic microwave background shift parameter. In the
aforementioned investigations, a flat background (Minkowski space-time) had
been adopted despite the fact that according to Visser, \textit{
\textquotedblleft in a cosmological setting it is no longer obvious that we
should use the Minkowski metric as a background"}. If a flat background is
adopted, then the only issue resulting in an expanding dynamic metric is to
assume that the divergence of the massive graviton tensor is a source for
the divergence of the matter stress-energy tensor [27]. In Visser's
theory the divergence of the graviton tensor is set to zero and,
consequently, when the dynamical equations are linearized around the
background metric, the Hilbert-Lorentz condition appears naturally and not
as a gauge one. It is worth mentioning that other bimetric gravity theories
with generally a flat prior geometry have been elaborated in the past, in
particular the approach by Rosen or the vector-bimetric theory by Rastall
(see, for instance, [32,33] for reviews). These theories are quite
distinct since the lagrangean density from which the field equations are
derived differs drastically from that proposed by Visser.

In the present paper, the Visser's theory is revisited as well as
applications to cosmology. We will show that the graviton tensor derived by
Visser [26] must be corrected by a factor equal to the ratio between the
background and dynamical metric determinants. The conservation laws are also
revisited and the following question is examined: if the divergence of the
graviton tensor is equal to zero (as it should be expected from general
arguments based on a variational approach of gravity) is it possible to find
an adequate background metric from which a cosmology compatible with the
present data emerges? A positive answer can be given but the solutions able
to mimic the $\Lambda$CDM cosmology are not completely satisfactory, since
they require that the background metric tensor be proportional to the scale
factor describing the dynamics of the universe. This paper is organized as
follows: in Section II the Visser's theory is revisited and the correct
graviton tensor is derived. An alternative definition of the lagrangean
density describing the graviton field is given as well as the resulting
field equations. In Section III an application of the theory to cosmology is
discussed and finally, in Section IV the main conclusions are presented.

\bigskip

\noindent \textbf{II. THE FIELD EQUATIONS\ AND\ THE CONSERVATION\ LAWS}

\bigskip

In this paper, the notation $m_{\alpha \beta }\equiv diag\left(
-1,+1,+1,+1\right) $ (Minkowski metric in cartesian coordinates) will be
adopted as well as units such $G=c=\hbar =1$. The action proposed by Visser
[26] reads
\begin{equation}
S=\int d^{4}x\left\{\frac{1}{16\pi}\left[\sqrt{-g}R+\sqrt{-b}
L_{mass}\left(g,b\right)\right] +\sqrt{-g}L_{matter}\left(g,X\right)
\right\}   \label{visserlagr}
\end{equation}
where $b_{\alpha \beta}$ represents the background metric tensor, 
$g_{\alpha\beta}$ the dynamical metric tensor and $X$ stands for any non
gravitational field. Both metrics are required to have a lorentzian
signature $(-,+,+,+)$. The lagrangean of the massive graviton is given
explicitly by 
\begin{equation}
L_{mass}\left( g,b\right) =-\frac{1}{4}m^{2}\left\{ b^{\alpha \beta }b^{\mu
\nu }\left( g_{\alpha \mu }-b_{\alpha \mu }\right) \left( g_{\beta \nu
}-b_{\beta \nu }\right) -\frac{1}{2}\left[ b^{\alpha \beta }\left( g_{\alpha
\beta }-b_{\alpha \beta }\right) \right] ^{2}\right\} .  \label{gravitonlagr}
\end{equation}%
The contravariant tensor $b^{\alpha \beta}$ is defined from $b_{\alpha\beta}$ 
by the inversion relation $b_{\alpha\beta }b^{\beta\gamma}=\delta _{\alpha }^{\gamma}$, while 
$g^{\alpha\beta}$ is defined from
the relation $g_{\alpha\beta}g^{\beta\gamma }=\delta _{\alpha}^{\gamma }$
as usually. The coefficient $-1/4$ ensures that in the weak field
limit, i.e., when $b_{\alpha \beta }=m_{\alpha \beta }$ and $g_{\alpha \beta
}=m_{\alpha \beta }+h_{\alpha \beta }$ with $\left\vert h_{\alpha \beta
}\right\vert <<1$, the graviton field in vacuum obeys the Klein-Gordon
equation. Since the construction of the action defined in eq.~\ref%
{visserlagr} is mainly motivated by such a requirement, it is worth pointing
that any alternative action leading to the same linearized equations would
be acceptable a priori. In particular, the graviton lagrangean density can
be defined in terms of the dynamical metric, i.e., $\sqrt{-g}L_{mass}$
instead of $\sqrt{-b}L_{mass}$. This corresponds to an alternative to
Visser's proposal that admits the same usual weak field limit. Such an
alternative for the graviton lagrangean density will be examined in some
more detail at the end of this section.

\bigskip

\noindent \textbf{A. Conservation laws}

\bigskip

From the variation of the action (eq.~\ref{visserlagr}) with respect to the
field $X$ (here, for simplicity, we assume the presence of only one
non-gravitational field) one obtains the Lagrange equation describing the
dynamics of the considered field. Since the matter lagrangean density 
$L_{matter}$ depends only on the field $X$ and on the dynamical metric tensor
(the field $X$ couples with $g_{\alpha \beta }$ only), then the
diffeomorphism invariance of $L_{matter}$ leads immediately to a
conservation equation expressed by the null covariant divergence condition
\begin{equation}
\nabla _{\alpha}T^{\alpha\beta }=0  
\label{matterconserv}
\end{equation}
where $T^{\alpha\beta}$ is the stress-energy tensor of matter [31]. If
the condition above is ignored, some inconsistencies may appear in the
physical laws describing the dynamics of non-gravitational fields. It is
worth recalling that the null divergence of the Einstein's tensor leads only
to the null divergence of the sum of tensors constituting the right side of
the field equations derived by varying the complete action of the theory (in
our case eq.~\ref{visserlagr}) with respect to the metric tensor $g_{\alpha\beta}$. If different 
stress-energy tensors are present, the null
divergence of the Einstein tensor is not equivalent to eq.~\ref{matterconserv}. In this case 
the validity of eq.~\ref{matterconserv} implies an additional condition to be fulfilled by the
solutions of the field equations.

\bigskip

\noindent \textbf{B. Field equations}

\bigskip

Varying the action defined by eq.~\ref{visserlagr} with respect to the
metric tensor $g_{\alpha \beta }$ leads to
\begin{equation}
G^{\alpha \beta}=8\pi T^{\alpha\beta}+8\pi T_{mass}^{\alpha\beta}
\label{fieldeq}
\end{equation}
where $G^{\alpha\beta}=R^{\alpha\beta}-\frac{1}{2}Rg^{\alpha\beta}$ is
the Einstein tensor, $R_{\alpha\beta}$ is the Ricci tensor and $R=R_{\beta}^{\beta}$ is 
the Ricci scalar. The matter stress-energy tensor $T^{\alpha\beta}$ is, as usually, defined by
\begin{equation}
T^{\alpha\beta }=\frac{2}{\sqrt{-g}}\frac{\delta\left(\sqrt{-g}
L_{mat}\right)}{\delta g^{\alpha\beta}}
\end{equation}
and the so-called graviton tensor by
\begin{equation}
T_{mass}^{\alpha \beta}=-\frac{m^{2}}{16\pi }\frac{\sqrt{-b}}{\sqrt{-g}}
\left(b^{\mu\alpha}b^{\nu\beta}-\frac{1}{2}b^{\mu\nu}b^{\alpha\beta}
\right)\left(g_{\mu\nu}-b_{\mu\nu}\right)  
\label{gravistress}
\end{equation}
From the null divergence of Einstein's tensor and eq. \ref{matterconserv},
it results
\begin{equation}
\nabla _{\alpha}T_{mass}^{\alpha\beta }=0.  
\label{gravitconserv}
\end{equation}
It should be emphasized that both conservation laws expressed by eqs.~\ref{matterconserv} and
\ref{gravitconserv} must be satisfied and these are necessary
conditions to be taken into account when considering solutions of the field
equations. From these equations, it is trivial to verify that Visser's
theory reduces to GR when the graviton mass $m$ vanishes. As we shall see in
details in appendix A, a class of solutions in which the dynamical metric
tensor is proportional to that of the background metric exists, which can be
generated from GR solutions including a cosmological constant.

\bigskip

\noindent \textbf{C. An alternative lagrangean density}

\bigskip

As already mentioned, the requirement that the graviton field obeys the
Klein-Gordon equation in the weak field (or linear) limit does not fix
the form of the lagrangean density. An alternative to Visser's proposal is
to consider a lagrangean density defined in terms of the dynamical metric
tensor, i.e., $\sqrt{-g}L_{mass}$ and, in this case, the total action of the
theory is
\begin{equation}
S=\int d^{4}x\sqrt{-g}\left\{ \frac{1}{16\pi }\left[ R+L_{mass}\left(
g,b\right) \right] +L_{matter}\left( g,X\right) \right\} 
\end{equation}
The resulting field equations are essentially identical to eq.~\ref{fieldeq}
but now the graviton tensor reads 
\begin{eqnarray}
T_{mass}^{\alpha\beta}=-\frac{m^2}{16\pi}\left[b^{\mu\alpha}b^{\nu\beta}g_{\mu\nu}+b^{\alpha\beta}
\left(1-\frac{1}{2}b^{\mu\nu}g_{\mu\nu}\right)\right] \\   \nonumber
-\frac{m^2}{16\pi} g^{\alpha\beta}\left[\frac{1}{4}b^{\mu\sigma}b^{\nu\rho}g_{\mu \nu}g_{\sigma\rho}
-\frac{1}{8}\left(b^{\mu\nu}g_{\mu\nu}\right)^{2}+\frac{1}{2}b^{\mu \nu}g_{\mu\nu }-1\right] 
\end{eqnarray}

It should be emphasized that as in the case of the original Visser's
proposal, the field equations derived from this alternative lagrangean
density admit solutions generated from GR equations including a cosmological
constant (see details in appendix A).

\bigskip

\noindent \textbf{III - FRW COSMOLOGY}

\bigskip

Recently, different authors have considered the Visser's theory to describe
the dynamics of the universe ([27-30]), claiming
that the graviton mass term appearing in the Friedman equations is able to
drive the observed acceleration of the expansion of the universe. In these
investigations, a Minkowski spacetime was assumed as a background and, in
this case, as a consequence of eq.~\ref{gravitconserv}, it results that the
scale factor must be constant or, in other words, the universe must be
\textquotedblleft static". In order to maintain a flat and static background
(Minkowski), the aforementioned authors assumed that the conservation law
expressed by eq.~\ref{gravitconserv} is violated and, in order to satisfy
the condition $\nabla _{\alpha }G_{\beta }^{\alpha }=0$, they have
hypothesized that the divergence of the graviton tensor is a (negative)
source term for the divergence of the stress-energy tensor of the matter.

Here the FRW cosmology is revisited in the context of the Visser's theory
but with preservation of the conservation laws as discussed in the previous
Section. Considering the standard FRW form for the dynamical metric we have
\begin{equation}
ds^{2}=-dt^{2}+a\left(t\right)^{2}\left[d\chi^{2}+F_{k}\left(\chi
\right)^{2}\left(d\theta^{2}+\sin^{2}\theta d\varphi^{2}\right)\right] 
\label{RWdynmetric}
\end{equation}
where $k=+1,0,-1$ corresponds respectively to spherical, flat or hyperbolic
spatial sections and, accordingly, $F_{k}\left(\chi\right) =\sin\chi
,\chi ,\sinh\chi$. Following Visser ([26]), we will search possible
non-static solutions for the background metric, i.e., 
\begin{equation}
d\sigma ^{2}=-B\left(t\right)^{2}dt^{2}+A\left(t\right)^{2}\left[d\chi^{2}
+G_{k}\left(\chi\right)^{2}\left(d\theta^{2}+\sin^{2}\theta d\varphi^{2}\right)\right]   
\label{RWbackmetric0}
\end{equation}
where $A$ and $B$ are positive functions of the cosmic time. From the
spatial components of eq.~\ref{gravitconserv} we have necessarily $G_{k}(\chi)=F_{k}(\chi)$. The 
function $A\left(t\right)$ may be
interpreted as a "background scale factor" while a \textquotedblleft
background cosmic time" may be defined by $\int B\left( t\right) dt$.
Matter, as usually, is described as a perfect fluid with energy density $\varepsilon\left(t\right)$ 
and pressure $P\left(t\right)$, linked by an equation of state $P=P\left(\varepsilon\right)$.

Besides the time component of matter conservation (eq.~\ref{matterconserv}),
the other equations describing the dynamics of the universe are the
components (0~0) and (1~1) of the field eq.~\ref{fieldeq}, since the remaining
components do not provide any additional information. On the other side, a
relation between the coefficients of the background and of the dynamical
metric can be obtained from eq.~\ref{gravitconserv}. It is useful to define
the functions $\Xi\left(t\right)=A\left(t\right)/a\left(t\right)$ and 
$\Psi\left(t\right)=B\left(t\right)/\Xi\left(t\right)$, which should
be positive since $A$, $B$ and the scale factor $a$ are positive functions.
In terms of these new functions the background metric can be expressed as 
\begin{equation}
d\sigma^{2}=\Xi\left(t\right)^{2}\left\{-\Psi\left(t\right)^{2}dt^{2}+a\left(t\right)^{2}
\left[d\chi^{2}+F_{k}\left(\chi\right)^{2}\left(d\theta ^{2}+\sin^{2}\theta d\varphi^{2}\right)\right]
\right\}   
\label{RWbackmetric}
\end{equation}

The usual procedure in the context of the Visser's theory is to define first
a background metric and then, from the field equations and conservation
laws, to derived the dynamics of the universe (supposing that the properties
of matter, like the equation of state are given a priori). In this case, any
cosmological scale parameter $a(t)$ can be obtained by an appropriate choice
of the background but this not necessarily implies that we have a physically
acceptable solution. Another possibility to tackle the problem is to search
for metric coefficients of the background spacetime once a dynamical
solution is imposed to the field equations or, in other words, to work in
the opposite route of the usual procedure. This approach will be adopted
here. From the field components (0~0) and (1~1) of eq.~\ref{fieldeq} we have
\begin{equation}
\left(\frac{\dot a}{a}\right)^2+\frac{k}{a^{2}}=\frac{8\pi }{3}\varepsilon +
\frac{m^{2}}{12}\Sigma  
\label{RW00}
\end{equation}
and
\begin{equation}
2\frac{\ddot a}{a}+\left(\frac{\dot a}{a}\right)^2+\frac{k}{a^{2}}
=-8\pi P+\frac{m^{2}}{4}\frac{(2\Psi ^{2}\Xi ^{2}-\Psi ^{2}-1)}{\Psi}
\label{RW11}
\end{equation}
where the notation $\dot{X}\equiv dX/dt$ was adopted (It is worth mentioning
that in above equations $\Xi $ and $\Psi $ are functions of the cosmic time $t$). The 
new function $\Sigma $ in eq.~\ref{RW00} is defined as
\begin{equation}
\Sigma =\frac{1}{\Psi^{3}}\left(1+2\Psi^{2}\Xi^{2}-3\Psi^{2}\right) 
\label{gravit00}
\end{equation}
The space components of eq.~\ref{gravitconserv} are trivially satisfied
while the time component provides a relation between the coefficients of the
background metric and the scale factor of the dynamical metric, namely
\begin{equation}
\frac{d\Sigma}{dt}+3\frac{\dot a}{a}\left(\Sigma +\frac{1+\Psi^{2}-2\Psi^{2}\Xi^{2}}{\Psi}\right) = 0  
\label{RWgravit1}
\end{equation}
or, equivalently
\begin{equation}
\frac{d\Sigma}{dt}=3\frac{\Psi^{2}-1}{\Psi^{3}}\left(1-\Psi^{2}+2\Psi^{2}\Xi^{2}\right)\frac{d\ln a}{dt}  
\label{RWgravit2}
\end{equation}
For the sake of completeness, we mention that the time component of eq.~\ref%
{matterconserv} can be written explicitly as%
\begin{equation}
\frac{d\left(\varepsilon a^{3}\right)}{dt}+3Pa^{2}\dot a = 0.
\label{RWmatter}
\end{equation}

Combining eqs.~\ref{RW00}, \ref{RW11} and \ref{gravit00} one obtains
\begin{equation}
\frac{\ddot a}{a}=-\frac{4\pi}{3}\left(\varepsilon+3P\right)+
\frac{m^{2}}{24\Psi^{3}}\left[2\Psi^{2}\Xi^{2}\left(3\Psi^{2}-1\right)
-\left(3\Psi^{4}+1\right)\right]   
\label{RWacc}
\end{equation}
The equation above shows that the last term on the right hand side
(depending on the graviton mass) can give a positive contribution to the
acceleration of the expansion of the universe. According to our adopted
approach, a solution for the background metric functions $\Xi (t)$ and $\Psi(t)$ can 
be obtained by adopting the following procedure: firstly, from the
matter equation of state and eq.~\ref{RWmatter}, the variation of the energy
density and of the pressure can be derived as a function of the scale
parameter $a$ or, equivalently, of the redshift $z$. Then, if the
acceleration parameter $q(z)=-{\ddot{a}}a/{\dot{a}}^{2}$ and the Hubble
parameter $H(z)$ are known from observations, the metric functions $\Xi (z)$
and $\Psi (z)$ can be determined from eqs.~\ref{RW00} and \ref{RWacc}.
Acceptable solutions require that the background metric functions be
positive and at the present time, in order to have a positive acceleration,
the following condition must be satisfied (including all the physical
constants) 
\begin{equation}
\frac{1}{\Psi^{3}}\left[2\Psi^{2}\Xi^{2}(3\Psi^{2}-1)-(3\Psi^{4}+1)
\right] > \frac{12\Omega _{m}H_{0}^{2}\hbar^{2}}{m^{2}c^{4}}
\end{equation}
where $\Omega _{m}$ is the present matter density parameter and the metric
functions $\Xi$ and $\Psi$ are taken at the present time. The other
symbols have their usual meaning.

Although static solutions have only an academic interest, it is worth
mentioning that these solutions necessarily implies also a static background
geometry and vice-versa. Hence, the $\Lambda $CDM cosmology cannot be
reproduced if a Minkowski background is adopted. This point will be
considered in some more detail in appendix B.

\bigskip

\noindent \textbf{A. Back to the\ }$\Lambda $\textbf{CDM\ cosmology}

\bigskip

The present observations (luminosity distance of type Ia supernovae, the
baryon acoustic peak (BAO) and the cosmic microwave background (CMB) shift
parameter) are quite well fitted by the so-called $\Lambda $CDM model. Is it
possible to mimic such a cosmological model within the framework of the
Visser's theory? A positive answer can be obtained if one identifies the
last term on the right side of eq.~\ref{RW00} as 
\begin{equation}
\frac{m^{2}}{12}\Sigma =\frac{1}{3}\Lambda 
\end{equation}
In this case the function $\Sigma$ must be a constant and eq.~\ref{RWgravit2} has two possible solutions. The 
first corresponds to $\Psi^{2}$=1 and $\Xi^{2}=1+2m^{-2}\Lambda$. Since $\Lambda >0$, we 
have $\Xi^{2}>1$. This particular solution corresponds to a class of solutions
discussed in appendix A, referred to the cosmological case. The second
possibility corresponds to $\Psi^{2}=(1-2\Xi^{2})^{-1}$ and $\Sigma =-2\sqrt{1-2\Xi^{2}}$. However, this 
solution implies a negative cosmological
constant, which is not supported by observations and will be not discussed
further.

\bigskip

\noindent \textbf{B. }$\Lambda $\textbf{CDM\ and the alternative Visser's
theory}

\bigskip

In the framework of the alternative definition of the graviton lagrangean
density, the $(0~0)$ component of the field equations is formally identical
to eq.~\ref{RW00} but now the function $\Sigma$ is defined as
\begin{equation}
\Sigma =\frac{5-\left(18-12\Xi^{2}\right)\Psi^{2}-\left(3-12\Xi^{2}+8\Xi^{4}
\right)\Psi^{4}}{4\Psi^{4}\Xi^{4}}
\end{equation}
Another equation relating the metric coefficients can be obtained from the
conservation law expressed by eq.~\ref{gravitconserv} whose time component is
\begin{equation}
\frac{d\Sigma}{dt}=3\frac{\Psi^{2}-1}{\Psi^{4}\Xi^{4}}\left(1-\Psi^{2}+2\Psi^{2}
\Xi^{2}\right)\frac{d\ln a}{dt}.
\end{equation}
Again, to mimic the $\Lambda$CDM cosmology we require that $\Sigma$=constant and, in 
this case, two solutions are possible: $\Psi^{2}$=1 and $\Psi^{2}=(1-2\Xi^{2})^{-1}$ 
(Notice that the solution $\Psi^{2}$=1 corresponds to a case discussed in appendix A). Contrary to the solutions
derived from equations based on the original Visser proposal, here the two
solutions are compatible with a positive cosmological constant and a lower
limit for the gravition mass can be derived from the observed value of $\Lambda$. In order to have 
at least one solution with a positive cosmological constant, the mass of the graviton must 
be higher than a critical value given by
\begin{equation}
m>2\sqrt{3\Omega _{\Lambda }}\frac{\hbar H_0}{c^{2}}\sim 7.7\times 10^{-66}\ g
\end{equation}
where we have adopted $\Omega _{\Lambda }=0.7$ and $H_0=70kms^{-1}Mpc^{-1}$
.

\bigskip

\noindent \textbf{IV - CONCLUSIONS}

\bigskip

In the present investigation the Visser's theory was revisited and, in
particular, we have found that the original graviton tensor must be
corrected by a factor equal to the ratio between the square root of the
determinant of the background metric and that of the dynamical metric. This
correction is not relevant when the usual weak field approximation is
considered but is of fundamental importance in cosmological applications. We
have also considered an alternative to the graviton lagrangean density
proposed by Visser and the consequent modifications in the graviton tensor
but both approaches lead to the same linear field equations.

We have also shown that the field equations of the theory when combined to
the conservation laws are able to mimic a $\Lambda $CDM cosmology if and
only if the background metric functions are proportional to the scale factor
defining the dynamical metric. This is in contradiction with claims in the
literature based on investigations considering a Minkowski background and an
abandon of the conservation laws expressed either by eq.~\ref{matterconserv}
or eq.~\ref{gravitconserv}. However, the results are somewhat disappointing
since only a particular form of the background metric, whose choice was
based on a priori cosmological considerations leads to satisfactory results.

Since the background required for compatibility between the theory and
cosmological observations turns to be time dependent, a natural question
raises: how this affects locally physical process occurring in the linear
regime? As argued by Visser, in the usual weak field approximation the
background metric should be Minkowiskian and, in fact, an expanding
background metric can be put in the form $m_{\alpha \beta }$ by a suitable
choice of coordinates if the relevant timescales are orders of magnitude
smaller than the cosmological timescale. This is the case when planetary
motions or the propagation of light within scales of the order of the solar
system are considered. However if this is true at a given instant, this may
not be the case (say) some billions years later. In this situation,
performing an expansion of the dynamical metric a Klein-Gordon equation is
not obtained. These considerations may lead into specific cosmological
signatures on the local dynamics, opening eventual new constraints on the
graviton mass by local observations. These aspects will be analyzed in a
forthcoming paper.

\bigskip

\bigskip

\noindent\lbrack 1] A.G. Riess et al., AJ \textbf{116}, 1009 (1998)

\noindent\lbrack 2] S. Perlmutter et al., ApJ \textbf{517}, 565 (1999)

\noindent\lbrack 3] D. Larson et al., ApJS \textbf{192}, 16 (2011)

\noindent\lbrack 4] E. Bianchi and C. Rovelli, arXiv:1002.3966v3

\noindent\lbrack 5] A. de Roany and J.A. de Freitas Pacheco,
Gen.Rel.Grav. \textbf{43}, 61 (2011)

\noindent\lbrack 6] O.F. Piattella, J.C. Fabris and W. Zimdahl,
arXiv:1103.1328

\noindent\lbrack 7] R. P. Woodard, Lect. Notes Phys. \textbf{720}, 403
(2007)

\noindent\lbrack 8] A. de Felice and S. Tsujikawa, \textit{f(R) Theories}
in Living Reviews in Relativity, www.livingreviews.org/Irr-2010-3

\noindent\lbrack 9] S. Tsujikawa, Mod. Phys. Lett. \textbf{25}, 843 (2010)

\noindent\lbrack 10] S. del Campo and P. Salgado, Class.Quant. Grav. 
\textbf{20}, 4331 (2003)

\noindent\lbrack 11] Y. Fujii and K. Maeda, \textit{The Scalar-Tensor
Theory of Gravitation}, Cambridge University Press, (2003)

\noindent\lbrack 12] V. Faraoni, \textit{Cosmology in Scalar-Tensor Gravity}%
, Kluwer Academic Publishers (2004)

\noindent\lbrack 13] G. Barenboim and J. D. Lykken, JCAP \textbf{3}, 17
(2008)

\noindent\lbrack 14] S. Unnikrishnan and T. R. Seshadri, IJMPD \textbf{17}%
, 2007 (2008)

\noindent\lbrack 15] N. Banerjee and K. Ganguly, IJMPD \textbf{18}, 445
(2009)

\noindent\lbrack 16] C. E. M. Batista and W. Zimdahl, Phys. Rev. \textbf{%
D82}, 023527 (2010)

\noindent\lbrack 17] M. Park, K. M. Zurek and S. Watson, Phys. Rev. 
\textbf{D81}, 124008 (2010)

\noindent\lbrack 18] M. Fierz and W. Pauli, Proc.R.Soc. \textbf{A173}, 211
(1939)

\noindent\lbrack 19] H. van Dam and M.J.G. Veltman, Nucl.Phys. \textbf{B22}%
, 397 (1970)

\noindent\lbrack 20] V.I. Zakharov, JETP Lett \textbf{12}, 312 (1970)

\noindent\lbrack 21] A.I. Vainshtein, Phys.Lett. \textbf{B39}, 393 (1972)

\noindent\lbrack 22] T. Damour, I.I. Kogan and A. Papazoglou, Phys.Rev. 
\textbf{D67}, 064009 (2003)

\noindent\lbrack 23] G.R. Dvali, G. Gabadadze and M. Porrati, Phys.Lett. 
\textbf{B485}, 208 (2000)

\noindent\lbrack 24] E. Babichev, C. Deffayet and R. Ziour, Phys.Rev. 
\textbf{D82}, 104008 (2010)

\noindent\lbrack 25] C. de Rham and G. Gabadadze, Phys. Rev. \textbf{D82},
044020 (2010)

\noindent\lbrack 26] M. Visser, Gen. Rel. Grav. \textbf{30}, 1717 (1998)

\noindent\lbrack 27] M.E.S. Alves, O.D. Miranda and J.C.N. de Araujo,
Gen. Rel. Grav. \textbf{40}, 765 (2008)

\noindent\lbrack 28] M.E.S. Alves, F.C. Carvalho, J.C.N. de Araujo,
O.D. Miranda, C.A. Wuensche and E.M. Santos, Phys. Rev. \textbf{D82}, 023505
(2010)

\noindent\lbrack 29] S. Basilakos, M. Plionis, M.E.S. Alves and J.A.S.
Lima, Phys. Rev. \textbf{D83}, 103506 (2011)

\noindent\lbrack 30] H.W. Lee, K.Y. Kim and Y.S. Myung, arXiv:1106.6114

\noindent\lbrack 31] L.D. Landau and E.M.Lifshitz, \textit{The Classical
Theory of Fields} Addison-Wesley, Reading, Massachusetts, Pergamon, London
(1971)

\noindent\lbrack 32] C. M. Will, \textit{Theory and Experiment in
Gravitational Physics}, Cambridge University Press (1993)

\noindent\lbrack 33] C. M. Will, \textit{The Confrontation Between General
Relativity and Experiments}, in Living Reviews in Relativity,
www.livingreviews.org/Irr-2006-3

\bigskip

\bigskip

\noindent \textbf{APPENDIX A - }$\Lambda$\textbf{GR solutions vs Visser's
theory}

\bigskip

For a given background metric $b_{\mu \nu}$, let us search for solutions in
which the dynamical metric $g_{\mu \nu}$ is proportional to the background
metric tensor, i.e.,
\begin{equation}
g_{\mu \nu }=\lambda ^{-2}b_{\mu \nu}  
\label{dynproptoback}
\end{equation}
where $\lambda^{2}$ is a positive constant. In this case, from eqs.~\ref{gravistress} 
(see text) one obtains

\begin{equation}
T_{mass}^{\alpha \beta }=\frac{1-\lambda ^{2}}{16\pi }m^{2}g^{\alpha \beta }
\end{equation}
Consequently the field equation takes formally the following structure
\begin{equation}
G^{\alpha \beta}+\frac{\lambda^{2}-1}{2}m^{2}g^{\alpha \beta }=8\pi
T^{\alpha\beta}
\end{equation}
Notice that the above equation satisfies consistently eq.~\ref{gravitconserv}
and indicates that all GR solutions including a cosmological constant $\Lambda $ are also 
solutions of Visser's massive gravity with a background
metric tensor
\begin{equation}
b_{\alpha\beta}=\left(1+\frac{2\Lambda}{m^{2}}\right)g_{\alpha\beta}
\label{visser_vs_LGR}
\end{equation}
This result indicates that solutions based on eq.~\ref{dynproptoback} are
possible only if the background metric tensor is proportional to a solution
of Einstein equations including a cosmological constant and with the same
matter stress-energy tensor.

Similarly, if the alternative lagrangean density for the graviton field is
considered, one obtains for the graviton tensor 
\begin{equation}
T_{mass}^{\alpha\beta }=\frac{1}{16\pi}\left(\frac{1}{\lambda^{2}}-1
\right)\left(\frac{2}{\lambda^{2}}-1\right)m^{2}g^{\alpha\beta}
\end{equation}
that replaced in the field equation leads to
\begin{equation}
G^{\alpha\beta}+\left(1-\frac{1}{\lambda^{2}}\right)\left(\frac{1}{\lambda^{2}}-\frac{1}{2}
\right) m^{2}g^{\alpha\beta}=8\pi T^{\alpha\beta}
\end{equation}
Thus, as in Visser's original theory, all GR solutions including a
cosmological constant term are also solutions of the massive gravity field
equations but now the proportional constant is given by
\begin{equation}
\left(1-\lambda^{-2}\right)\left(2\lambda^{-2}-1\right)=\frac{2\Lambda}{m^{2}}  
\label{visser2_vs_LGR}
\end{equation}

\bigskip

\noindent \textbf{APPENDIX\ B : Static cases in Visser's theory}

\bigskip

Static solutions are those with a constant scale factor $a$ (or $\dot{a}=0$). The 
equations to be considered in this case are eqs.\ref{RW00}, \ref{RW11}
and \ref{RWgravit1} (see text). From these equations one obtains
\begin{equation}
\frac{k}{a^{2}}=\frac{8\pi}{3}\varepsilon +\frac{1}{12}m^{2}\Sigma
\end{equation}
where
\begin{equation}
\Sigma =\frac{1}{\Psi^{3}}\left[1+2\Psi^{2}\Xi^{2}-3\Psi^{2}\right]
\end{equation}
\begin{equation}
\frac{k}{a^{2}}=-8\pi P+\frac{1}{4}m^{2}\frac{1}{\Psi}\left[2\Psi ^{2}\Xi^{2}-\Psi^{2}-1\right]
\end{equation}
with $\dot{\Sigma}=0$ (static case), which implies that $\Sigma =$constant.
The constancy of the energy density and of the pressure can be derived from
the first equation and the equation of state. Thus, both $\Sigma$ and the
quantity $\Psi^{-1}\left[1+\Psi^{2}-2\Psi^{2}\Xi^{2}\right]$ are
constants, indicating that the background metric is necessarily static.

For a given $\varepsilon$, $P$, $k$ and $a$ (the case $P=0$ corresponds to a
dust filled static universe in the context of Visser's theory), the metric
coefficient $\Psi $ is obtained by solving the equation
\begin{equation}
2\left( \frac{3k}{a^{2}}-8\pi \varepsilon \right)\Psi^{3}-m^{2}\Psi^{2}-2
\left(\frac{k}{a^{2}}+8\pi P\right)\Psi+m^{2}=0
\end{equation}
and, consequently, $\Xi$ satisfies
\begin{equation}
\Psi^{2}\Xi^{2}=\frac{\Psi^{2}+1}{2}-\frac{2}{m^{2}}\left(\frac{k}{a^{2}}+8\pi P\right)\Psi 
\end{equation}
The equations above permit to obtain the metric coefficients $\Psi $ and $\Xi$.

Reciprocally, if we have a static background, we have necessarily $A$ and $B$
or $\Psi\Xi $ and $\Xi a$ constants (Notice that these imply also that $a/\Psi$ is a constant). Inserting 
these relations into eq.~\ref{RWgravit2}
and after a straightforward calculation one obtains
\begin{equation}
\left( 2\Psi^{2}\Xi^{2}-\Psi^{2}+1\right)\frac{d\Psi}{dt}=0.
\end{equation}
Thus, one has either $2\Psi^{2}\Xi^{2}-\Psi^{2}+1=0$ or $d\Psi/dt=0$.
Both possibilities imply $\Psi$=constant and $a$= constant, i.e., a
static universe.

\end{document}